\documentclass[achemso,twocolumn,superscriptaddress,floatfix,amsmath,amssymb]{revtex4-2}
\usepackage{amsmath}
\usepackage{amssymb}
\usepackage{braket}
\usepackage{graphicx}
\usepackage[table]{xcolor}
\usepackage{verbatim}
\usepackage{float}
\usepackage{color}
\usepackage{siunitx}
\usepackage{gensymb}
\usepackage{bm}
\usepackage{xcolor}
\usepackage{multirow}
\usepackage{footnote}
\usepackage{bbm}
\usepackage{enumitem}
\usepackage{oplotsymbl}
\usepackage{bbding}

\usepackage{array}
\newcolumntype{P}[1]{>{\centering\arraybackslash}p{#1}}
\newcolumntype{Q}[1]{>{\raggedleft\arraybackslash}p{#1}}
\newcolumntype{R}[1]{>{\raggedright\arraybackslash}p{#1}}

\newcommand{\rr}{\mathbf{r}}

\newcommand{\bvec}[1]{\mathbf{\boldsymbol{#1}}}

\begin{document}
\title{Competition of moir{\'e} network sites to form electronic quantum dots in reconstructed MoX\textsubscript{2}/WX\textsubscript{2} heterostructures}

\author{Isaac Soltero}
\email{isaac.solteroochoa@manchester.ac.uk}
\affiliation{Department of Physics and Astronomy, University of Manchester. Oxford Road, Manchester, M13 9PL, United Kingdom}
\affiliation{National Graphene Institute, University of Manchester. Booth St.\ E., Manchester, M13 9PL, United Kingdom}
\author{Mikhail A. Kaliteevski}
\affiliation{Department of Physics and Astronomy, University of Manchester. Oxford Road, Manchester, M13 9PL, United Kingdom}
\affiliation{National Graphene Institute, University of Manchester. Booth St.\ E., Manchester, M13 9PL, United Kingdom}
\author{James G. McHugh}
\affiliation{Department of Physics and Astronomy, University of Manchester. Oxford Road, Manchester, M13 9PL, United Kingdom}
\affiliation{National Graphene Institute, University of Manchester. Booth St.\ E., Manchester, M13 9PL, United Kingdom}
\author{\\Vladimir V. Enaldiev}
\affiliation{Department of Physics and Astronomy, University of Manchester. Oxford Road, Manchester, M13 9PL, United Kingdom}
\affiliation{National Graphene Institute, University of Manchester. Booth St.\ E., Manchester, M13 9PL, United Kingdom}
\author{Vladimir I. Fal'ko}
\email{vladimir.falko@manchester.ac.uk}
\affiliation{Department of Physics and Astronomy, University of Manchester. Oxford Road, Manchester, M13 9PL, United Kingdom}
\affiliation{National Graphene Institute, University of Manchester. Booth St.\ E., Manchester, M13 9PL, United Kingdom}
\affiliation{Henry Royce Institute for Advanced Materials, University of Manchester. Oxford Road, Manchester, M13 9PL, United Kingdom}

\begin{abstract}

Twisted bilayers of two-dimensional semiconductors offer a versatile platform to engineer quantum states for charge carriers using moir{\'e} superlattice effects. Among the systems of recent interest are twistronic MoSe${}_{2}$/WSe${}_{2}$ and MoS${}_{2}$/WS${}_{2}$ heterostructures, which undergo reconstruction into preferential stacking domains and highly strained domain wall networks, determining the electron/hole localization across moir{\'e} superlattices. Here, we present a catalogue of options for the formation of self-organized quantum dots and wires in lattice-reconstructed marginally twisted MoSe${}_{2}$/WSe${}_{2}$ and MoS${}_{2}$/WS${}_{2}$ bilayers, fine tuned by the twist angle between the monolayers from perfect alignment to $\theta \sim 1^{\circ}$, and by choosing parallel or anti-parallel orientation of their unit cells. The proposed scenarios of the quantum dots and wires formation are found using multi-scale modelling that takes into account the features of strain textures caused by twirling of domain wall networks.

\end{abstract}

\maketitle

Among various heterostructures of two-dimensional materials, bilayers of same-chalcogen transition metal dichalcogenides (TMDs) ${\rm MoX}_{2}/{\rm WX}_{2}$ (X$=$Se, S) stand out because of a uniquely small mismatch monolayers lattice constants ($\sim 0.3\%$ for MoSe${}_{2}$/WSe${}_{2}$ and $\sim 0.2\%$ for MoS${}_{2}$/WS${}_{2}$). An almost lattice matching at the interface promotes reconstruction of the moir{\'e} superlattice (mSL) in a bilayer into domains where monolayer lattices conform towards each other, separated by domain wall networks (DWNs) which absorb hydrostatic and shear (for small angle twisted bilayers) strain. Such reconstruction was observed in bilayers both assembled by 2D crystal transfer \cite{mcgilly2020visualization,rosenberger2020twist,sung2020broken,shabani2021deep,weston2020atomic,Rupp_2023} and synthesised using CVD \cite{li2023lattice}. For bilayers with a larger twist, lattice reconstruction is less prominent, however, those also feature moir{\'e} superstructures, which were extensively investigated in the context of localization of charge carriers and interlayer excitons in specific stacking regions of the mSL \cite{rivera2015observation,jin2019observation,tran2019evidence,seyler2019signatures,zhao2023excitons,fang2023localization}.

\begin{figure}[t!]
    \centering
    \includegraphics[width=1.0\columnwidth]{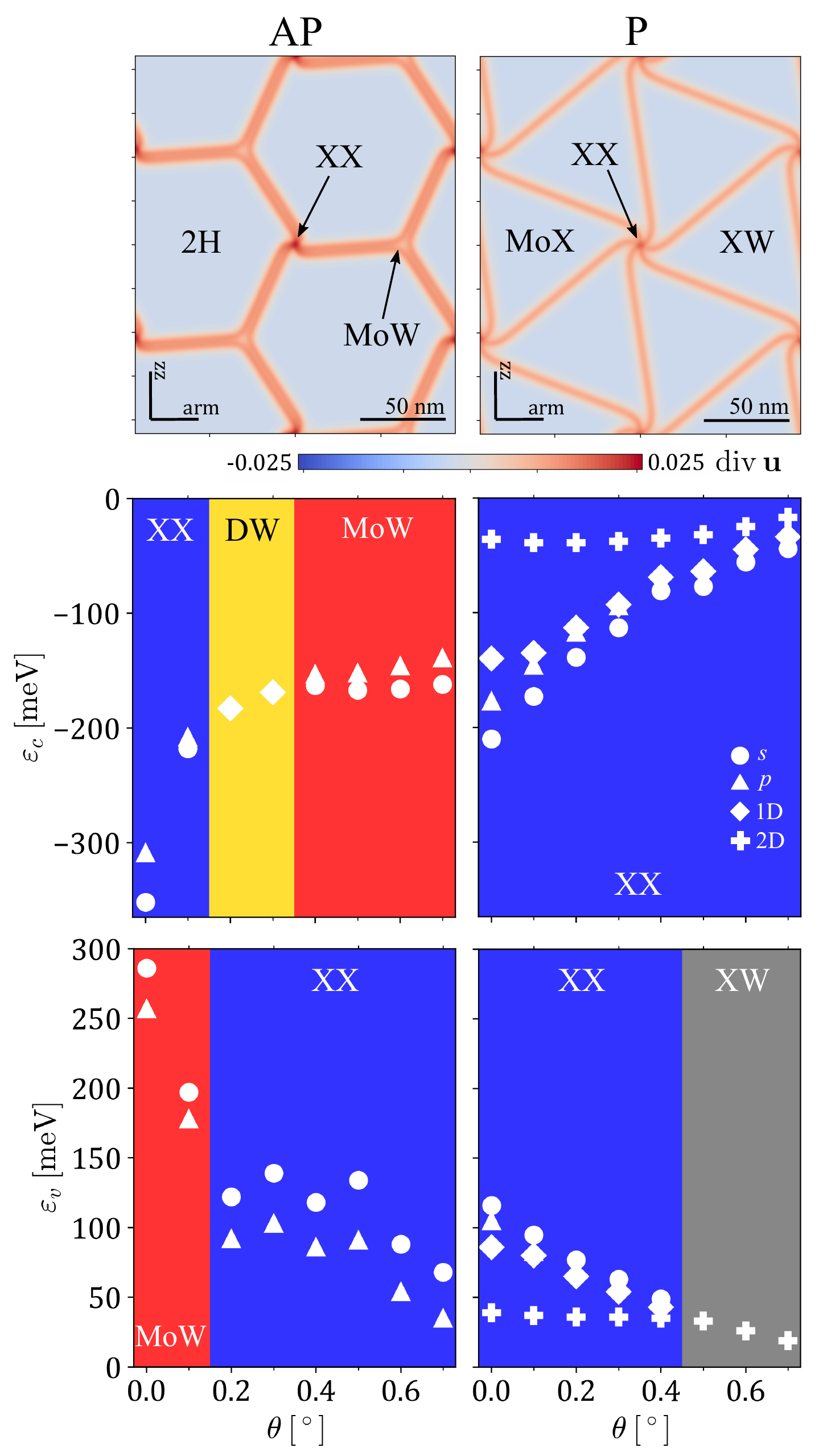}
    \caption{Hydrostatic strain maps (top panel) in the Mo layer for AP- (left) and P-${\rm MoSe}_{2}/{\rm WSe}_{2}$ (right) with $\theta=0^{\circ}$. Stacking configurations for domains, and the zigzag (zz) and armchair (arm) crystallographic axes of the monolayers are indicated in each map. Bottom panel shows where on the reconstructed bilayer domain structure we expect the localization of electrons ($\varepsilon_{c}$) and holes ($\varepsilon_{v}$). We use color coding to demonstrate the crossover in the position of a quantum dot for a conduction band electron or a valence band hole as a function of twist angle. White symbols indicate energies for quantum dots ($s$ and $p$),  wires along domain walls (1D) and XW domain boxes (2D).}
    \label{fig:Divergence_Intro}
\end{figure}

In terms of electronic properties of ${\rm MoX}_{2}/{\rm WX}_{2}$, which are type-II heterostructures \cite{gong2013band}, the mapping \cite{shabani2021deep,ferreira2021band,enaldiev2021piezoelectric} of conduction/valence band edge variation across the mSL appears to be sensitive to the strain developed upon bilayer's lattice reconstruction. In particular, for marginally twisted bilayers, the formation of DWNs absorbs the dilation/compression and torsion required for adjusting the two crystalline lattices within domains (both in terms of the lattice mismatch of ${\rm MoX}_{2}$ and ${\rm WX}_{2}$ and an interlayer twist) \cite{mcgilly2020visualization,rosenberger2020twist,sung2020broken,shabani2021deep,weston2020atomic,Rupp_2023,li2023lattice}. Typical structures are illustrated in top panels of Fig. \ref{fig:Divergence_Intro} by the maps of hydrostatic strain in MoSe${}_{2}$: honeycomb for bilayers with antiparallel (AP) orientation of unit cells and triangular for parallel (P) orientation. For AP bilayers, hexagonal domains correspond to 2H stacking, while corners of the DWN feature energetically unfavorable XX (chalcogen over chalcogen) and MoW (metal over metal) stackings \cite{carr2018relaxation,enaldiev2020stacking,enaldiev2021piezoelectric}. P bilayers feature MoX (metal over chalcogen) and XW (metal under chalcogen) stacking domains, with XX stacking DWN nodes \cite{carr2018relaxation,enaldiev2020stacking,enaldiev2021piezoelectric}.

\begin{figure}
    \centering
    \includegraphics[width=0.95\columnwidth]{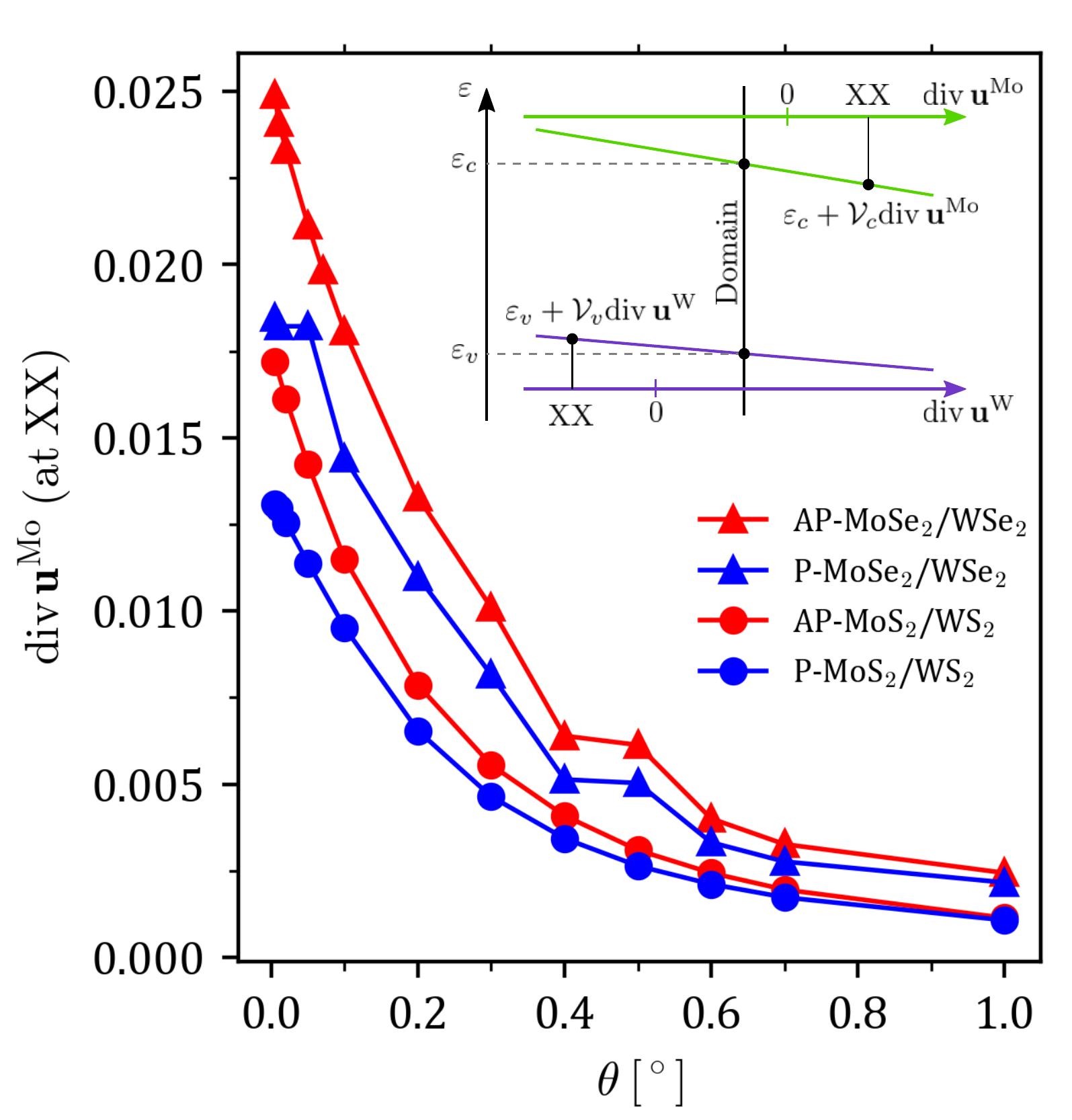}
    \caption{Twist angle dependence of hydrostatic strain in the Mo layer, ${\rm div}\, \bvec{u}^{\rm Mo}$, at the XX nodes of AP- and P-${\rm MoX}_{2}/{\rm WX}_{2}$ (${\rm X}={\rm Se, S}$) bilayers. Inset shows linear shift of $K$-valley conduction (green) and valence (purple) band edges in the MoX${}_{2}$ and WX${}_{2}$ layers, respectively. Solid black line indicates the value of hydrostatic strain in individual layers inside stacking domains.}
    \label{fig:Divergence}
\end{figure}

In this paper, we demonstrate a high sensitivity of the band edge properties of marginally twisted (small angle $|\theta|< 1^{\circ}$) ${\rm MoX}_{2}/{\rm WX}_{2}$ bilayers to the lattice reconstruction and DWN formation. Recently, it has been noted that, while the nodes of such DWNs represent the areas of the largest strain, both hydrostatic (lattice dilation in MoX${}_{2}$ \textit{vs} compression in WX${}_{2}$) and shear \cite{naik2018ultraflatbands,enaldiev2020stacking,weston2020atomic,enaldiev2022self}, the  higher energetic cost of hydrostatic strain can be reduced by transforming some dilation/compression into shear achieved via `twirling' of DWN \cite{dai2016twisted,kaliteevsky2023twirling,maity2021reconstruction,mesple2023giant}, as marked on the maps in Fig. \ref{fig:Divergence_Intro}. Here, we highlight the role of hydrostatic strain, because it causes stronger band edge shifts than shear deformations \cite{conley2013bandgap,dhakal2017local,zollner2019strain,enaldiev2022self} (see inset in Fig. \ref{fig:Divergence}), hence, affecting electron and hole confinement across the reconstructed moir{\'e} structure. Highest values for hydrostatic strain (corresponding to XX nodes), quantified in Fig. \ref{fig:Divergence} as the strain tensor trace, $u_{ii}\equiv {\rm div}\, \bvec{u}$, reveal its predominant role for nearly aligned heterobilayers. 

The resulting variety is illustrated for ${\rm MoSe}_{2}/{\rm WSe}_{2}$ by the bottom panels in Fig. \ref{fig:Divergence_Intro}. In particular, for an almost perfectly aligned AP bilayer ($\theta<0.15^{\circ}$), corresponding to that ($\theta[{\rm rad}]<\delta\approx 3\times 10^{-3}$, a relative lattice mismatch between MoX${}_{2}$ and WX${}_{2}$), we find that electrons and holes are localized at quantum dots at the opposite corners of hexagonal DWNs, with binding energies of both $s$- and $p$-like orbitals in the range of 200$-$300 meV, as shown in Fig. \ref{fig:Divergence_Intro}. For slightly higher twist angles, $\theta>0.15^{\circ}$, we find that the area of confinement for valence band holes relocate from MoW to XX corners of the DWN, whereas for the conduction band electrons the quantum dot at the XX node blends into the surrounding stretches of domain walls for an intermediate range of angles ($0.15<\theta<0.35^{\circ}$), which, now, can be considered as quantum wires. Further increasing the twist angle ($\theta>0.35^{\circ}$), electrons eventually locate at the MoW cornes of the DWN. 

This contrasts very different systematics emerging for P-${\rm MoSe}_{2}/{\rm WSe}_{2}$, where, for $\theta<0.45^{\circ}$, the lowest energy states for both electrons and holes appear to be at the XX nodes of the triangular DWN network. The right hand side panels in Fig. \ref{fig:Divergence_Intro} indicate that, while for the larger angles the `XX quantum dots' persist for the conduction band electrons (bound states only 10$-$20 meV below the lowest 1D band inside the domain wall `wires'), the valence band holes get deconfined into the XW stacking domain areas, where the band edge is promoted by the interlayer charge transfer \cite{enaldiev2021piezoelectric,wang2022interfacial,ferreira2021weak,weston2022interfacial,rogee2022ferroelectricity,deb2022cumulative}, as compared to MoX domains. As discussed in detail in the rest of the paper, similar scenarios also appear in ${\rm MoS}_{2}/{\rm WS}_{2}$.

\textit{Methods}. The multi-scale modelling approach implemented in this study (see SI) consists of the following steps:
\begin{enumerate}[label=(\roman*)]
    \item To describe the variation of the local stacking across the moir{\'e} pattern, we use a model developed in Refs. \onlinecite{enaldiev2021piezoelectric,enaldiev2020stacking} which combines the microscopically computed (using DFT) stacking dependent adhesion energies between the ${\rm MoX}_{2}$ and ${\rm WX}_{2}$ crystals with elasticity theory description of intralayer deformations.
    \item The computed deformation fields in each layer are used to describe energies of electron (in ${\rm MoX}_{2}$) and hole (in ${\rm WX}_{2}$) band edges at the $K$ points of the TMD Brillouin zone. To mention, both hydrostatic and shear deformations have opposite signs in ${\rm MoX}_{2}$ and ${\rm WX}_{2}$), and in twirled DWNs feature less hydrostatic strain as compared to geometrically straight network structures \cite{carr2018relaxation,enaldiev2022self}. 
    \item In the analysis of band energies we account for both the band edge shift due to hydrostatic strain \cite{conley2013bandgap,dhakal2017local,zollner2019strain,enaldiev2022self} and piezocharges due to shear strain, both computed locally across the domain network and plotted as 2D maps over the moir{\'e} supercell to identify the potential confinement areas for charge carriers near the DWN nodes.
    \item Having identified those confinement areas we use an effective mass approximation for electrons/holes to compute their respective quantum dot spectra.
\end{enumerate}

The lattice structure of the reconstructed bilayer is described employing 2D displacement fields $\bvec{u}^{({\rm Mo/W})}(\rr)$, which determine the local lateral offset between the two crystals,
\begin{equation}\label{Eq:r0}
    \rr_{0}(\rr) = \delta \rr + \theta\hat{\bvec{\mathrm{z}}}\times \rr + \bvec{u}^{\rm Mo}(\rr) - \bvec{u}^{\rm W}(\rr).
\end{equation}
Here, $\delta\equiv 1- a_{\rm W}/a_{\rm Mo}$ is a relative lattice mismatch between ${\rm MoX}_{2}$ and ${\rm WX}_{2}$, defined in terms of their respective lattice parameters $a_{\rm Mo}$ and $a_{\rm W}$. Note that for each local stacking configuration we use the optimal interlayer distance, $d(\rr_{0})$ \cite{enaldiev2020stacking} (that is, corresponding to the minimal of adhesion energy for each $\rr_{0}$). Then, we minimise the total energy of the bilayer over the moire supercell, taking into account both elastic and adhesion energy contributions (see SI for details). Two representative solutions for strains in P and AP bilayers are displayed in top panels of Fig. \ref{fig:Divergence_Intro} as color maps of the developed hydrostatic strain in ${\rm MoX}_{2}$ layer, ${\rm div}\,\bvec{u}^{\rm Mo}$. A small hydrostatic strain inside domains compensates small lattice mismatch between the layers (${\rm div}\,\bvec{u}^{\rm Mo}\approx -\delta$ and ${\rm div}\,\bvec{u}^{\rm W}\approx \delta$); the corresponding strain energy in both layers is fully compensated by the gain in adhesion energy \cite{enaldiev2020stacking}, as proven by the observation of thermal annealing MoSe${}_{2}$/WSe${}_{2}$ bilayers into fully commensurate heterostructures \cite{baek2023thermally}. We emphasize that only hBN encapsulated TMD heterobilayers are considered in this analysis, which allows to neglect the formation of bulges and out-of-plane defects.

To include hydrostatic strain in the conduction and valence band edge shift analysis, we performed DFT calculations using the Quantum ESPRESSO code \cite{giannozzi2009quantum} (see SI). In all cases, band edge shifts can be approximated by $\mathcal{V}_{c/v}\,{\rm div}\,\bvec{u}^{\rm Mo/W}$ with the DFT-computed values $\mathcal{V}_{c}^{{\rm MoSe}_{2}}=-11.57$ eV, $\mathcal{V}_{v}^{{\rm WSe}_{2}}=-5.76$ eV, $\mathcal{V}_{c}^{{\rm MoS}_{2}}=-12.45$ eV, and $\mathcal{V}_{v}^{{\rm WS}_{2}}=-5.94$ eV, which are in agreement with several previous \textit{ab initio} studies \cite{conley2013bandgap,dhakal2017local,zollner2019strain,enaldiev2022self}. This hydrostatic strain effect is incorporated in the band energy profiles (see SI for details),
\begin{equation}\label{BandEdge}
    \delta\varepsilon_{c/v}(\rr) = \mathcal{V}_{c/v}\,{\rm div}\,\bvec{u}^{\rm Mo/W}(\rr) - e\varphi_{c/v}(\rr) \mp \frac{1}{2}\Delta(\rr),
\end{equation}
where $\varphi_{c/v}(\rr)$ is a strain-induced piezoelectric potential (calculated for an hBN encapsulated bilayer), and $\Delta(\rr)$ is a stacking-dependent energy shift due to a weak interlayer polarization (largely varying across the reconstructed moir{\'e} supercell in P and weakly in AP bilayers \cite{enaldiev2021piezoelectric}). This determines the effective quantum dot profiles for the conduction and valence bands, implemented as databases for heterostructures with different twist angles, all of which have a $C_{3}$ symmetry for both P and AP alignment. Then, we use brute force diagonalization to analyze the confinement of electrons and holes (see SI for details).

\begin{figure*}
    \centering
    \includegraphics[width=1.0\textwidth]{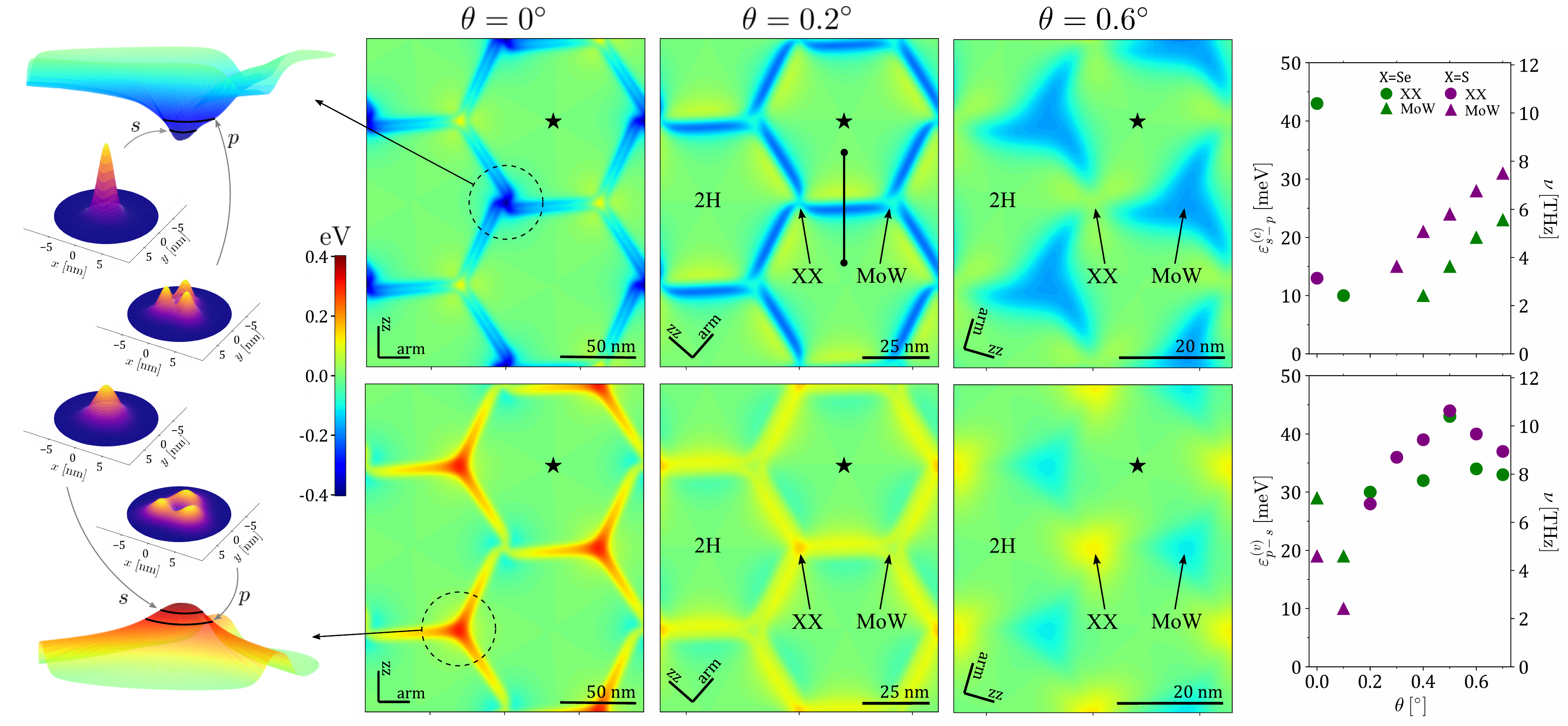}
    \caption{Formation of quantum dots and wires in marginally twisted AP-MoSe${}_{2}$/WSe${}_{2}$. Middle panel presents the modulation of conduction (top) and valence (bottom) band edges for different twist angles. All energies are calculated from the corresponding $c$/$v$ band edge energy in the middle of 2H stacking domains (marked by stars), and each map is rescaled according to its corresponding moir{\'e} periodicity (scale bar and crystallographic axes of the monolayers are indicated in all maps). The segment in the intermediate conduction band edge map indicates the one-dimensional potential profile for the calculation of bound states in domain walls. Left panel shows the conduction and valence quantum dot profiles for $\theta=0^{\circ}$, with squared moduli of $s$ and $p$ states indicated in each case. Right panel shows the twist angle dependence of the energy of intralayer $s$-$p$ transitions, $\varepsilon_{s-p}$, in quantum dots (and their corresponding frequency, $\nu$) in AP-${\rm MoX}_{2}/{\rm WX}_{2}$ (X=Se, S) structures.
    }
    \label{fig:BandEdgesAP}
\end{figure*}

\textbf{Band edges in AP-MoX${}_{2}$/WX${}_{2}$}. In this section we discuss bound states in AP-${\rm MoX}_{2}/{\rm WX}_{2}$ (X$=$Se, S), where the inner areas of hexagonal shape domains are occupied by a 2H-type stacking (top layer metals over bottom layer chalcogens, and vice versa). The conduction and valence band edge profiles, calculated with respect to the center of the 2H stacking domains, are shown in Fig. \ref{fig:BandEdgesAP}. Below, we discuss separately the electron and hole confinement.

\textit{Quantum dots and wires at conduction band edges (electrons).} For closely aligned AP-${\rm MoSe}_{2}/{\rm WSe}_{2}$ structures ($\theta<0.15^{\circ}$), the joint effect of hydrostatic strain and piezoelectric potential results in deep quantum wells at the XX nodes, forming quantum dot states in those regions, with the energies of bound $s$- and $p$-states plotted in Fig. \ref{fig:Divergence_Intro} with circles and triangles, respectively. In an intermediate range of angles, $0.15^{\circ}<\theta<0.35^{\circ}$, the reduction of hydrostatic strain, and the inversion of the sign of piezoelectric charge (see SI) results in the disappearance of the local energy minima in the DWN nodes (see Fig. \ref{fig:BandEdgesAP}), pushing conduction band edge into domain walls, where 1D bound states can be viewed as stretches of quantum wires. Lowest subband energies for such quantum wires are shown in Fig. \ref{fig:Divergence_Intro} with diamonds. For higher twist angles, $\theta>0.35^{\circ}$, piezopotential plays the predominant role, producing quantum dots in the MoW corners, with the energies of bound states represented as circles and triangles in Fig. \ref{fig:Divergence_Intro}.

We note that approaching the interval $0.15^{\circ}<\theta<0.35^{\circ}$ with the band edge forming domain wall quantum wires, $s$ and $p$ quantum dot bound states in XX ($\theta\approx0.1^{\circ}$) and MoW ($\theta\approx0.35^{\circ}$) become nearly degenerate. However, further increase of the twist angle promotes MoW quantum dots, leading to a lift of the degeneracy and reappearance of $s$ and $p$ bound states.

\textit{Quantum dots in MoW and XX corners at the valence band edge (holes).} For the valence band in a bilayer with $\theta[{\rm rad}]<\delta$, the piezopotential contribution to the band edge energy negates the confining effect of hydrostatic strain at the XX nodes. Instead, it produces distinct quantum dots in the MoW areas (see Fig. \ref{fig:BandEdgesAP}) with binding energies of $s$- and $p$-states shown in Fig. \ref{fig:Divergence_Intro} with circles and triangles, respectively. The reduction of hydrostatic strain for $\theta[{\rm rad}]>\delta$ leaves the piezopotential as the dominant effect. Therefore, the change of piezocharges sign at $\theta[{\rm rad}]\approx \delta$ shifts the quantum dot position towards the XX nodes, where the binding energies (of holes) are smaller.

The scenario of electron and hole confinement by DWNs in marginally twisted AP-${\rm MoS}_{2}/{\rm WS}_{2}$ is similar (see SI). The only qualitative difference is that the electron confinement at XX nodes is fragile against electrons leaking into domain wall (DW) quantum wires. 

The overall evolution of the band edge landscape across moir{\'e} supercell is shown in Fig. \ref{fig:BandEdgesAP} for both conduction and valence band electrons, with a typical structure orbitals of $s$ and $p$ states attributed to the quantum dots. While the binding energies of such quantum dots states are plotted in Fig. \ref{fig:Divergence_Intro} (see SI for binding energies in AP-${\rm MoS}_{2}/{\rm WS}_{2}$), two right hand side panels in Fig. \ref{fig:BandEdgesAP} quantify  the excitation spectra of intradot intraband $s$-$p$ transitions for the confined electrons which are optically active in the THz range.

\begin{figure*}
    \centering
    \includegraphics[width=1.0\textwidth]{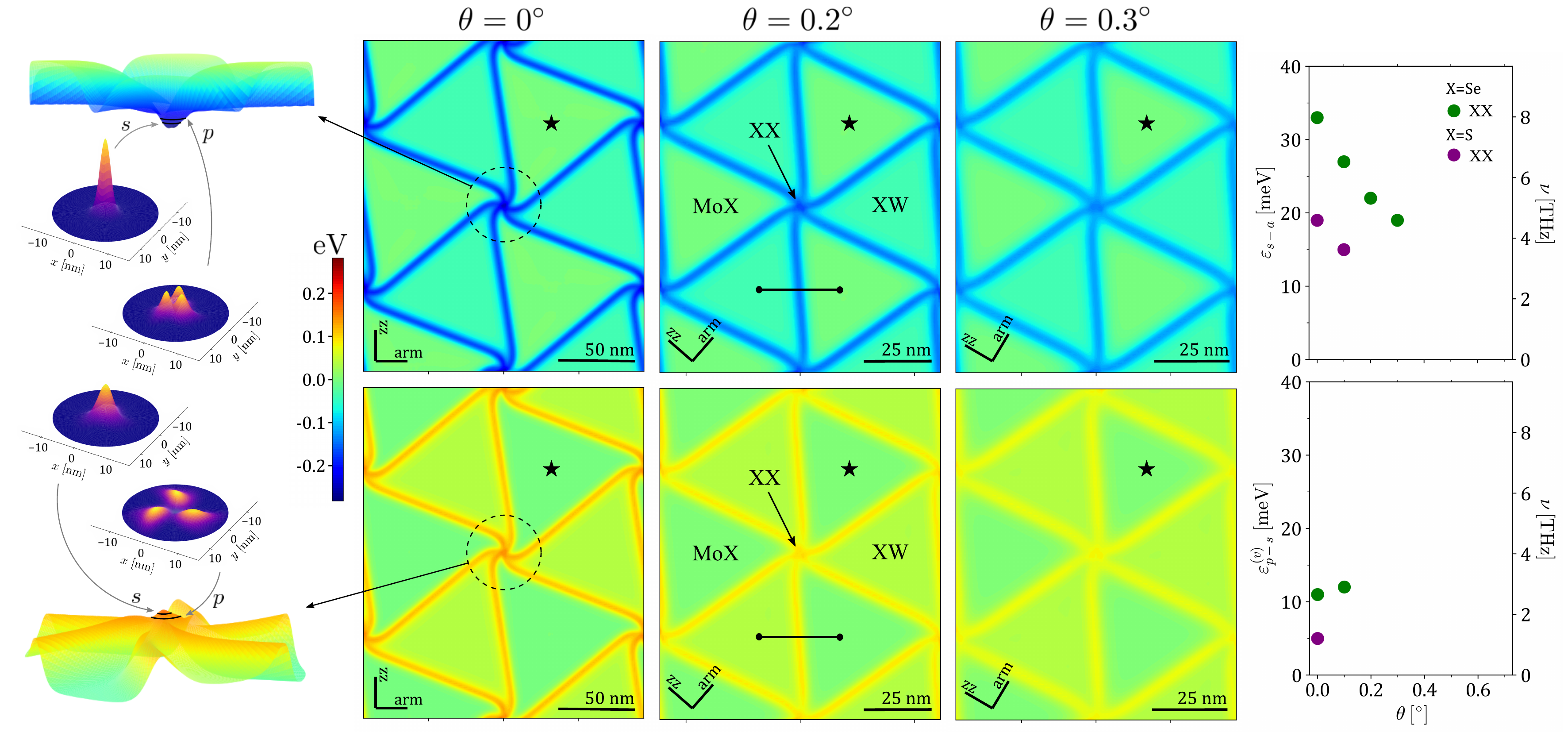}
    \caption{Formation of quantum dots, wires and 2D boxes in marginally twisted P-MoSe${}_{2}$/WSe${}_{2}$. Middle panel presents the modulation of conduction (top) and valence (bottom) band edges for different twist angles. All energies are calculated from the corresponding $c$/$v$ band edge energy in the middle of MoX stacking domains (marked by stars), and each map is rescaled according to its corresponding moir{\'e} periodicity (scale bar and crystallographic axes of the monolayers are indicated in all maps). The segments in the intermediate maps indicate the one-dimensional potential profiles for the calculation of bound states in domain walls. Left panel shows the conduction and valence quantum dot profiles for $\theta=0^{\circ}$, with squared moduli of $s$ and $p$ states indicated in each case. Right panel shows the twist angle dependence of the energy of intralayer $s$-$p$ transitions, $\varepsilon_{s-p}$, in quantum dots (and their corresponing frequency, $\nu$) in P-${\rm MoX}_{2}/{\rm WX}_{2}$ (X=Se, S) structures.}
    \label{fig:BandEdgesP}
\end{figure*}
 
\textbf{Band edges in P-MoX${}_{2}$/WX${}_{2}$}. In this section we discuss bound states in P-${\rm MoX}_{2}/{\rm WX}_{2}$ (X$=$Se, S), where DWNs separate triangular domains with MoX (top layer metals over bottom layer chalcogens) and XW (top layer chalcogens over bottom layer metals) stackings. The obtained band edge profiles are shown in Fig. \ref{fig:BandEdgesP}, where energies for the conduction (valence) band were calculated with respect to the highest (lowest) energy stacking domain, which corresponds to the MoX configuration. For highly aligned P-${\rm MoSe}_{2}/{\rm WSe}_{2}$, both conduction and valence band edge maps have their minimum value at the DWN intersections (XX nodes), with constant depth channels along domain walls, generated mainly by hydrostatic strain. Moreover, the energy difference between stacking domains, introduced by interlayer charge transfer, produces triangular quantum boxes in the XW stacking domains. As the twist angle is increased, there is a progressive reduction of the energy minima and the depth of the channels compared to the energy in stacking domains. This is driven by the drop in ${\rm div}\, \bvec{u}$, which is the only contribution along the DWN (piezopotential and charge transfer contributions are distributed over the stacking domains, see SI). Due to the difference in the magnitude of hydrostatic strain effects on the conduction and valence bands, below, we discuss electron and hole confinement separately.

\textit{Quantum dots and wires at conduction band edges (electrons).} In small angle P-${\rm MoSe}_{2}/{\rm WSe}_{2}$ ($\theta<0.35^{\circ}$), the strong effect of hydrostatic strain leads to the coexistence of quantum dots, wires and 2D boxes, whose bound state energies are shown in Fig. \ref{fig:Divergence_Intro}. Quantum dots at XX nodes hold bound states for the full range of twist angles studied, although $p$-states only persist up to $\theta=0.3^{\circ}$, spreading along the DWN for larger twist angles.

\textit{Quantum dots and wires at valence band edges (holes).} Despite having a qualitatively identical structure to conduction band profiles, valence band edges in P-${\rm MoSe}_{2}/{\rm WSe}_{2}$ acquire shallower confinement regions with quantum dot and wire bound states up to $\theta\approx0.45^{\circ}$ ($p$-states obtained only for $\theta<0.15^{\circ}$), with energies shown in Fig. \ref{fig:Divergence_Intro}. Increasing the twist angle ($\theta>0.45^{\circ}$) leads to relocalization of the lowest energy state for holes to the XW stacking domains. This crossover from quantum dots/wires regime to 2D confined states is described in Fig. \ref{fig:Divergence_Intro}.

Results for P-${\rm MoS}_{2}/{\rm WS}_{2}$ exhibit a slightly modified picture. In this case, electrons escape the DWN for $\theta>0.65^{\circ}$ (see SI) producing a weakly confined 2D state. Furthermore, holes reveal bound state energies only for a small range of angles ($\theta<0.25^{\circ}$), hosting $p$-states only for the aligned structures ($\theta=0^{\circ}$). For both ${\rm MoSe}_{2}/{\rm WSe}_{2}$ and ${\rm MoS}_{2}/{\rm WS}_{2}$ bilayers, the quantum dots in XX corners support $s$ and $p$ bound states only for almost aligned crystals, and in the two right hand side panels in Fig. \ref{fig:BandEdgesP} we quantify the energies of optically active intradot intraband $s$-$p$ transitions.

\textit{Discussion $\&$ conclusions}. Overall, the reported analysis of the band edge landscapes in lattice-reconstructed marginally twisted bilayers of same-chalcogen TMDs suggests an entertaining scenario that unfolds over a small range of misalignment angles, $|\theta|\leq 0.7^{\circ}$, for electrons'/holes' localization across the moir{\'e} supercell.

In particular, in perfectly aligned AP bilayers, we find distinct quantum dots in the opposite XX and MoW corners for electrons and holes, respectively, which swap upon the increase of twist angle beyond $\theta[{\rm rad}]>\delta$. This means that a very small variation in the heterostructure assembly conditions may qualitatively alter the nature of self-organized quantum dots in the reconstructed moir{\'e} pattern. Theoretically, the change of electron/hole quantum dot location as a function of a finely tuned twist angle can be considered as a crossover, and, specifically for electrons, such a crossover involves an intermediate regime where the conduction band edge passes through a quantum-wire-like domain wall. This results improves on a recently studied model \cite{enaldiev2022self} based on a domain wall structure without twirling, which largely overestimated the amount of hydrostatic strain, and therefore, the depth of the quantum dot profiles. For all these confinement regimes, we computed the binding energies for charge carriers (Fig. \ref{fig:Divergence_Intro} for AP-${\rm MoSe}_{2}/{\rm WSe}_{2}$ and SI for AP-${\rm MoS}_{2}/{\rm WS}_{2}$), taking into account the dielectric environment of hBN encapsulated bilayers (which is relevant for the piezopotential contribution towards band edge profiles), and quantified the energies of THz-active intradot transitions (Fig. \ref{fig:BandEdgesAP}). Additionally, the XX and MoW nodes feature peaks of strain-induced pseudomagnetic field \cite{enaldiev2020stacking,rostami2015theory}, which would split $\sigma_{+}$ and $\sigma_{-}$ polarized transitions for electrons/holes in $K$ and $K'$ valleys and we estimated to be of the order of $\sim 1$ meV. 

\begin{figure}
    \centering
    \includegraphics[width=1.0\columnwidth]{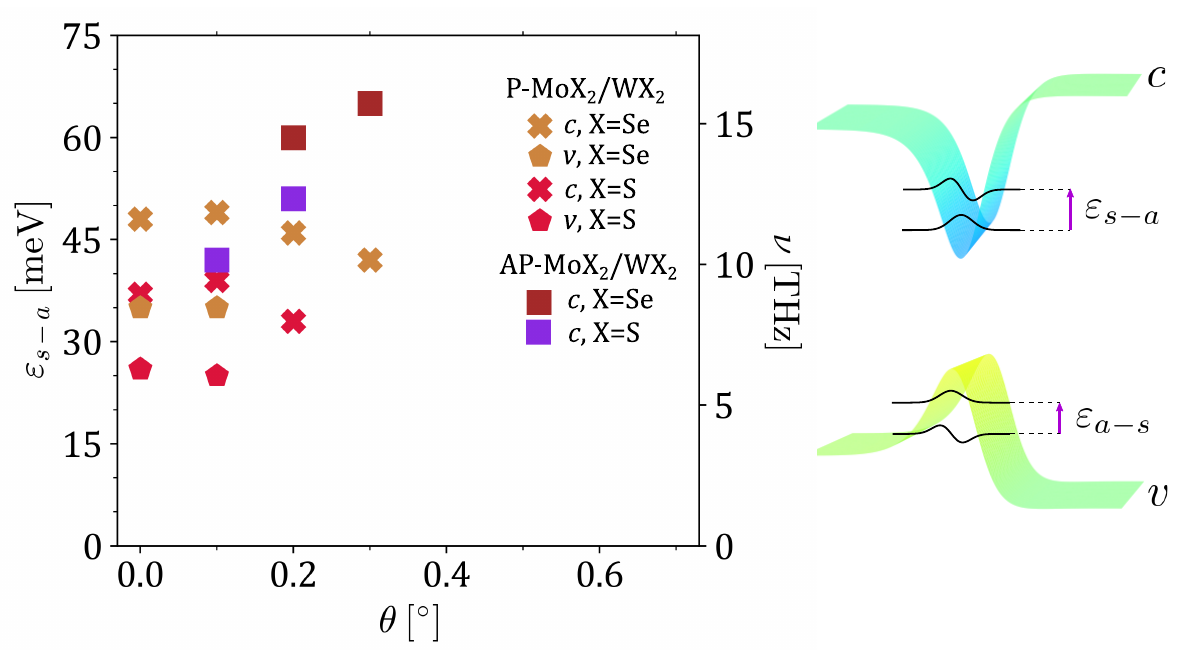}
    \caption{Twist angle dependence of transition energy (and their corresponding frequency) between the symmetric and anti-symmetric domain wall bound states in conduction  and valence bands of AP- and P-${\rm MoX}_{2}/{\rm WX}_{2}$. Right panel shows the energy profile for the conduction ($c$) and valence ($v$) one-dimensional channels in P-${\rm MoSe}_{2}/{\rm WSe}_{2}$ with $\theta=0^{\circ}$. The functions of the lowest energy states and the corresponding transition are schematized.
    }
    \label{fig:Transitions}
\end{figure}

In contrast, marginally twisted P bilayers feature quantum dots for both electrons and holes in the XX nodes of their DWN, though for holes such quantum dot confinement is quickly lost at twist angles $\theta[{\rm rad}]>\delta$ due to a carrier delocalization into triangular XW stacking domains (Figs. \ref{fig:Divergence_Intro} and \ref{fig:BandEdgesP}), accompanied by reduction of the $s$-$p$ on-dot orbital splitting (Fig. \ref{fig:BandEdgesP}). We also note that boundaries between MoX and XW domains in P bilayers may be considered as quantum wires, in particular for electrons. If such quantum wires are filled  by carriers due to doping, they would be characterized by subbands and THz-active intra-subbands transitions which energies are shown in Fig. \ref{fig:Transitions} (conduction and valence sub-bands transitions are marked by $\boldsymbol{\times}$ and $\pentagofill$, respectively). We note that such transition would be specific to P bilayers with small twist angles, as the second subband state in the wire would be lost much earlier than the lowest subband would delocalize into XW domains. For comparison, in Fig. \ref{fig:Transitions}, we also show the intersubband transition in the domain wall quantum wires (formed in AP bilayers in the above discussed  crossover regime and marked by $\blacksquare$), as notable on the band edge profile for electrons in $0.2^{\circ}$-twisted AP bilayer in Fig. \ref{fig:BandEdgesAP}.

While this analysis was focused on single electron or hole states, it also suggests the effect of DWN on few particle complexes. For example, the simultaneous confinement of both electrons and holes at the XX nodes of triangular DWNs in P bilayers (promoted by hydrostatic strain) points towards interlayer exciton localization at those nodes in bilayers with $\theta[{\rm rad}]<\delta$. This contrast to the charge (electron/hole) separation in the opposite DWN corners in AP bilayers across the same twist angle range. One may also speculate that the band edge profiles near the DWN in marginally twisted (both P and AP) bilayers would also confine X${}^{-}$ and X${}^{+}$ trions \cite{brotons2021moire,wang2021moire}, with the same scenario as for electrons and holes, respectively.

\textit{Acknowledgements}. The authors would like thank Marek Potemski and Fabio Ferreira for fruitful discussions. I.S. acknowledges financial support from the University of Manchester's Dean's Doctoral Scholarship. This work was supported by the EC-FET Core 3  European Graphene Flagship Project, EC-FET Quantum Flagship Project 2D-SIPC, EPSRC Grants EP/S030719/1 and EP/V007033/1, and the Lloyd Register Foundation Nanotechnology Grant.

\textit{Author contributions}. V.I.F. conceived the project. V.V.E. and M.A.K. developed the method for lattice reconstruction analysis, which was quantitatively implemented by I.S. and combined by I.S. and V.I.F. with numerical modelling of confined states in self-organized quantum dots. J.G.M. calculated microscopic parameters from \textit{ab initio} modelling. I.S. and V.I.F. wrote the paper in discussions with all authors.

\bibliography{References}

\end{document}